# Superconducting LaP$_2$H$_2$ with graphenelike phosphorus layers

Xing Li [1,*], Xiaohua Zhang [1,2,*], Aitor Bergara [3,4,5], Guoying Gao [6], Yong Liu [1], and Guochun Yang [1,2,†]

[1] *State Key Laboratory of Metastable Materials Science & Technology and Key Laboratory for Microstructural Material Physics of Hebei Province, School of Science, Yanshan University, Qinhuangdao 066004, China*
[2] *Centre for Advanced Optoelectronic Functional Materials Research and Key Laboratory for UV Light-Emitting Materials and Technology of Northeast Normal University, Changchun 130024, China*
[3] *Departamento de Física, Universidad del País Vasco-Euskal Herriko Unibertsitatea, UPV/EHU, 48080 Bilbao, Spain*
[4] *Donostia International Physics Center (DIPC), 20018 Donostia, Spain*
[5] *Centro de Física de Materiales CFM, Centro Mixto CSIC-UPV/EHU, 20018 Donostia, Spain*
[6] *Center for High Pressure Science (CHiPS), State Key Laboratory of Metastable Materials Science and Technology, Yanshan University, Qinhuangdao 066004, China*

Novel structural building blocks in compounds could induce interesting physical and chemical properties. Although phosphorus tends to form very different motifs, the existence of lone pair electrons has always prevented the formation of graphenelike structures. Here, the application of first-principles swarm structural calculations has allowed us to predict the stability of pressure-induced hexagonal LaP$_2$H$_2$ containing graphenelike phosphorus, which derives from the trigonal bipyramid configuration of P atoms regulated by symmetric hydrogen bonds. LaP$_2$ in LaP$_2$H$_2$ has the same configuration as MgB$_2$, and P and H atoms form a three-dimensional framework as H$_3$S. Interestingly, LaP$_2$H$_2$ shows a superconductivity dominated by the graphenelike phosphorus layer and its coupling with La atoms. On the other hand, LaP$_2$H$_2$ is not only superconducting at a lower pressure than the H-rich LaPH$_6$, but it also shows a superconducting transition temperature three times higher. Our work provides an example which extends the landscape of conventional superconductors at lower pressures.

The search for conventional superconductors has recently achieved a breakthrough, approaching the goal of room-temperature superconductivity [1–4]. Basic atomic building blocks in superconductors, such as hydrogen cages in LaH$_{10}$ [5] and pentagraphenelike hydrogen in HfH$_{10}$ [3], are decisive for this high-temperature superconductivity, which can be described by the Bardeen-Cooper-Schrieffer (BCS) theory [6], offering a clear guide for designing new superconducting materials [5,7,8]. Notably, first-principles calculations, combined with computational structural searches, play a key role in understanding the origin of superconductivity and in accelerating the discovery of superconductors [9–14].

MgB$_2$ became the milestone of BCS superconductors with an enhanced $T_c$ of 39 K [15], where B atoms adopt a graphenelike configuration with a $\sigma$ bonding in the boron plane, due to the $sp^2$ hybridization, and a delocalized $\pi$ bonding out of the boron plane, via the $p_z$ orbital hybridization [16]. Remarkably, the graphenelike boron layer induces a strong electron-phonon coupling (EPC), which is responsible for its superconducting transition [17,18].

Later, H$_3$S became the prototype of pressure-induced covalent hydride superconductors, displaying a remarkably high $T_c$ of 203 K at 155 GPa [1], reawakening the dream of room-temperature superconductivity [19,20]. H$_3$S shows an extraordinary structure, with H atoms symmetrically bonded to two S atoms in the body centered cubic sublattice, forming a three-dimensional (3D) framework [7]. The high-frequency H vibration mode induced by the covalent framework enhances the EPC responsible for superconductivity [21].

Additionally, the LaH$_{10}$ clathrate represented another kind of high-$T_c$ superconducting hydride, consisting of sodalitelike hydrogen cages [5] and a face centered cubic La sublattice stabilized by covalent and ionic interactions [22]. The H-derived high-frequency phonon mode strongly couples with La $f$ and H $s$ electronic states near the Fermi level ($E_F$) [23], yielding an exciting high $T_c$ of 260 K at 188 GPa [2].

On the other hand, it is well known that phosphorus (P) can form abundant allotropes [24] (e.g., white, black, and blue phosphorus) and diverse motifs [25] in its compounds (e.g., chains, rings, cages, and tubular frameworks). The appearance of nonplanar P configurations is one of their common characteristics, which is due to the repulsion between lone electron pairs and bonding electrons. In P-rich metal phosphides, P building units can be modulated by metallic elements, which is attributed to their different radii and electropositivities, and are closely related to superconductivity [26–28]. For instance, *Pnma* Nb$_2$P$_5$ with zigzag P chains shows a $T_c$ of 2.6 K at 0 GPa [26]; the $T_c$ of *I*4/*mmm* KP$_2$, consisting of puckered layers of P$_4$ rings, goes up to 22 K at 5 GPa [27], and *Cmmm* MgP$_2$, stabilized by a three-dimensional P covalent network, has a $T_c$ of 10.2 K at 150 GPa [28]. Therefore, the discovery of novel P motifs might provide an opportunity to find new superconductors with an enhanced $T_c$.

---

*These authors contributed equally to this work
†Corresponding author: yangguoc468@nenu.edu.cn

In addition, P is adjacent to S in the periodic table and both have similar electronegativities, thus, after the discovery of superconducting $H_3S$, P-H compounds were also considered as potential high-$T_c$ superconductors. First, Drozdov *et al.* [29] observed experimentally a $T_c$ of 103 K in compressed $PH_3$ at 207 GPa. However, its crystal structure and superconducting origin have not been unequivocally clarified. Several theoretical works found that all the considered P-H stoichiometries are thermodynamically metastable with respect to the decomposition into elemental P and solid $H_2$ [30–33]. Interestingly, some candidate phases have calculated $T_c$ values comparable to the measured ones (76 K for $C2/m$ $PH_2$ at 200 GPa [30], 78 K for $I4/mmm$ $PH_2$ at 220 GPa [31], and 83 K for $C2/m$ $PH_3$ at 200 GPa [32]). Unlike pressure-induced $H_3S$, metastability of $PH_3$ could be attributed to the fact that P has one valence electron ($3s^2 3p^3$) less than S ($3s^2 3p^4$).

Based on a physicochemical intuition, the insertion of metal atoms combined with variable pressures has become a strategy to improve the stability of the P-H system [34–36]. For example, several Li-P-H ternary compositions were predicted to be stable at high pressures [34,35]. Even more interestingly, two high-$T_c$ superconductors were proposed, $Pm$-3 $LiPH_6$ with a $T_c$ of 167 K at 200 GPa and $R$-3 $LiP_2H_{14}$ with 169 K at 230 GPa, where superconductivity is associated with a strong EPC of H-derived high-frequency vibrations, similar to $H_3S$. Later, $R$-3$m$ $TiPH_4$ was predicted to have a $T_c$ of 62 K at 250 GPa, closely related to the low-frequency vibrations of Ti and P atoms [36]. Until now, these studies focused on H-rich compositions. However, considering the fascinating characteristics of P, as mentioned above, there are expectations of stabilizing ternary compounds with novel P motifs by modulating the ratio of P and H, which may offer an opportunity to design new superconducting materials and understand the interplay of P motifs and H atoms on the superconducting mechanism.

La atom has a strong ability to donate electrons, stabilizing H cages and ternary hydrides, such as $LaN_2H_3$ [37] and $LaBH_8$ [37–39]. On the other hand, there appears puckered P layers composed of $P_{12}$ rings [40] in $LaP_5$ at ambient pressure. We cannot help but imagine whether introducing La atoms combined with pressure could result in the stable La-P-H compounds with novel P motifs and charming superconductivity. As expected, $LaP_2H_2$, consisting of a hitherto unknown graphenelike P configuration, becomes stable above 100 GPa. More interestingly, the graphenelike P layer and its coupling with La atoms are mainly responsible for the superconducting transition, analogous to the B layer in $MgB_2$. The $T_c$ value of 36.3 K at 125 GPa in $LaP_2H_2$ is much higher than the $T_c$ of 10.4 K at 200 GPa in H-rich $LaPH_6$.

Our structural search was performed by employing an intelligence-based particle-swarm optimization algorithm [41], as described in the Supplemental Material [49]. Reliable determination of the phase stability of ternary compounds requires knowledge on the stable structures of constituent elemental solids and binary compounds. The high-pressure phase diagrams of La-H [5,8,42,43] and P-H [30–33] systems, as well as their elemental solids (La [44,45], P [46], and $H_2$ [47]) are already known. However, stable La-P compositions at high pressures have so far rarely been explored [48]. Therefore, we first explored the stability of La-P compounds (Fig. S1 in the

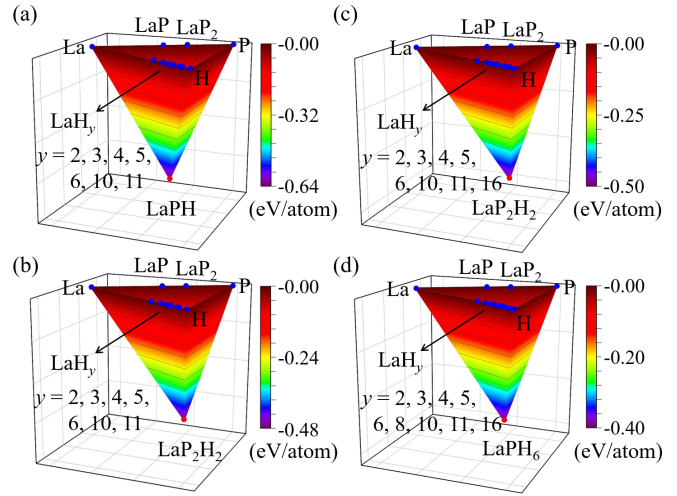

FIG. 1. Calculated stabilities of $LaP_xH_y$ derived from their formation enthalpies ($\Delta H$) relative to elemental solids and binary compounds at (a) and (b) 100 GPa, (c) 125 GPa, and (d) 150 GPa. To highlight the stability of predicted ternary compounds, the elements and stable binary phases are indicated by blue circles, and their $\Delta H$ values are projected onto the plane of 0 eV.

Supplemental Material [49]). The detailed structural characteristics and properties of stable binary La-P compounds will be reported elsewhere.

Taking into account both the high-$T_c$ H-rich compounds and the role of P motifs in the superconducting transition, we focused our structural search on P-rich and H-richer compounds. Energetic stabilities of a plethora of $LaP_xH_y$ ($x = 1$–3, $y = 1$–8) compounds are evaluated by their formation enthalpies relative to the dissociation into constituent elements and binary compounds at 100, 125, and 150 GPa. As shown in the convex hull diagrams (Fig. 1), three chemical compositions, LaPH, $LaP_2H_2$, and $LaPH_6$, become thermodynamically stable.

$LaP_2H_2$ crystallizes into a hexagonal structure (space group $P6/mmm$) at 100 GPa. LaPH and $LaPH_6$ stabilize into an orthorhombic structure with space groups $Pnma$ at 100 GPa and $Pmmn$ at 150 GPa, respectively. The three stable phases are dynamically stable as well, due to the absence of imaginary frequencies in their phonon dispersion curves (Fig. S2 [49]). Compared to Li-P-H and Ti-P-H systems [34–36], besides H-rich composition, P-rich $LaP_2H_2$ also emerges.

In addition to the difference in composition between La-P-H and Li/Ti-P-H [34–36], the three stable La-P-H compounds exhibit unique structural and bonding characteristics. Specifically, in $LaP_2H_2$ [Fig. 2(a)] P atoms form a desirable graphenelike configuration through an equivalent threefold coordination in the $ab$ plane [Fig. 2(b)]. The H atoms are symmetrically distributed between the two adjacent P planes, and each H atom bonds with two P atoms along the $c$ axis, forming a 3D P-H covalent framework. Thus, P atoms show a trigonal bipyramid configuration, as appeared in $PCl_5$ [61]. However, the $sp^2$ hybridization in the P plane and the two left electrons sitting in the $p_z$ orbitals that bond with two H atoms are in stark contrast with the $sp^3d$ hybridization in $PCl_5$. This is supported by the absence of a P $d$ orbital contribution and a strong overlap between P $p_z$ and H $s$ orbitals in the projected density of states (Fig. S3 [49]). The P-P and P-H bond lengths are 2.11 and 1.66 Å at 125 GPa, respectively, which are within the range of a covalent bonding [30–32,62], as is confirmed by the electron localization functional (ELF) [Fig. 2(c)]. In addition to the same arrangement of La and P atoms with respect to Mg and B, respectively, in MgB$_2$, each La atom coordinates with six H atoms.

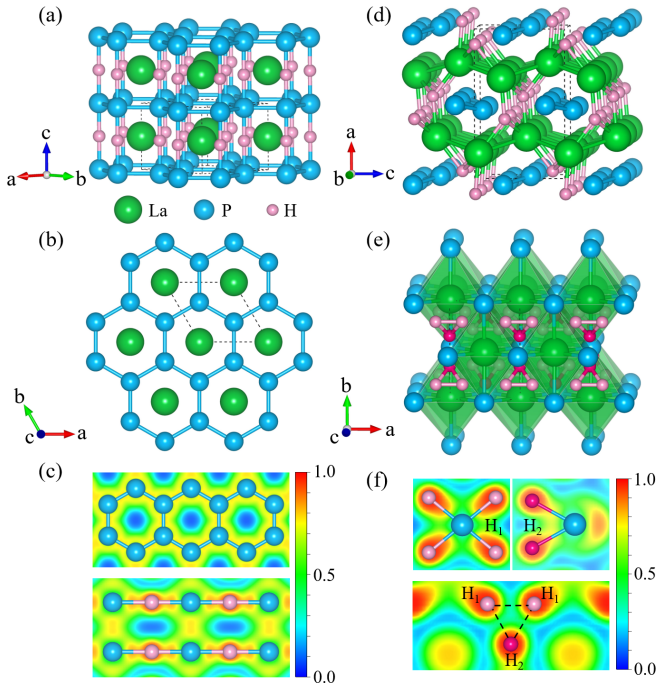

FIG. 2. The crystal structure of $P6/mmm$ LaP$_2$H$_2$ at 125 GPa in (a) side and (b) top view, (d) $Pnma$ LaPH at 100 GPa, and (e) $Pmmn$ LaPH$_6$ at 200 GPa. The ELF maps of (c) LaP$_2$H$_2$ on the (001) and (110) planes, and (f) LaPH$_6$.

On the other hand, the atomic arrangement of the symmetric P-H-P differs remarkably from those of binary P-H [30–32] and ternary Li/Ti-P-H compounds [34–36]. Symmetric hydrogen bonding has played an important role in raising the H vibration frequency and strengthening the EPC, as in H$_3$S [7], and BH [63] compounds. Such unique structure of LaP$_2$H$_2$ is stabilized by a charge transfer from La to P and H atoms, which facilitates the formation of symmetric hydrogen bonds. The Bader charge analysis [64] shows that each La atom loses 1$e^-$, and that both P and H atoms accept 0.3$e^-$ and 0.2$e^-$, respectively. Considering that a P atom just needs five valence electrons to form fivefold covalent bonds, the electrons donated by La atoms might occupy P-P antibonding orbitals under pressure, which is confirmed by the enhancement of localized electrons between P-P bonds after removing La atoms [Fig. S4(a)].

$Pnma$ LaPH [Fig. 2(d)] contains puckered La layers in the $bc$ plane, in which H atoms are located at the interlayer region, forming HLa$_4$ tetrahedra with La atoms in adjacent La layers. Interestingly, the HLa$_4$ tetrahedra are arranged alternately sharing edges along the $b$-axis direction, making spiral hexagonal channels. In the channels, P atoms show a zigzag-chain configuration with a P-P bond length of 2.12 Å. Bader charge analysis and the ELF map [Fig. S4(b)] indicate that P atoms accept electrons from La atoms and are covalently bonded via $sp^3$ hybridization. The shortest distance between P and H atoms is 2.12 Å, which is much longer than their covalent bond range [30–32], and means negligible P-H interaction.

In $Pmmn$ LaPH$_6$ [Fig. 2(e)], the edge-sharing distorted LaP$_6$ octahedra make a tetrahedron cavity to accommodate triangular H units. Each H triangle consists of two inequivalent H's, labeled as H$_1$ and H$_2$ atoms, with interatomic distances of 1.36 Å for H$_1$-H$_1$, and 1.35 Å for H$_1$-H$_2$. Such a unique H triangle is stabilized by strong P-H$_1$ and weak P-H$_2$ polar covalent bonds, with distances of 1.46 and 1.61 Å, respectively, as well as by a weak covalent interaction between H atoms [65,66], as identified by the ELF in Fig. 2(f). Apparently, these structural characteristics in LaPH$_6$ are distinct from the H-rich phases in Li/Ti-P-H systems, which usually contain PH$_y$ polyhedra [34–36].

Calculations of electronic properties clearly indicate that the three stable phases are metallic. Among them, LaP$_2$H$_2$ has the highest electronic density of states (DOS) at the $E_F$ at 125 GPa (Fig. S5). The projected electronic energy band of LaP$_2$H$_2$ [Fig. 3(d)] shows that the electronic states associated to La and P atoms are responsible for the metallicity, while H states are mainly located below the $E_F$ hybridized with P states, which stabilizes the P-H framework. A clear orbital overlap of P $s$, $p_x$, and $p_y$ states appears near the high symmetry point A (Fig. S5a), supporting the $sp^2$ hybridization. Additionally, there are steep bands along the Γ-A and A-H directions and a flat band around the Γ point near the $E_F$, corresponding to a large DOS.

To further analyze the electronic properties of LaP$_2$H$_2$, we also study the Fermi surfaces [Figs. 3(a)–3(c)] associated with the three bands that contribute significantly at the $E_F$. One of the Fermi surfaces shows a horn shape [Fig. 3(a)] and the other two have cylindrical features [Figs. 3(b) and 3(c)], which are quite similar to the topology of the Fermi surfaces in MgB$_2$ [16], confirming the similarities between LaP$_2$H$_2$ and MgB$_2$ (Figs. S7a–S7c). In detail, the horn-shaped Fermi surface in LaP$_2$H$_2$ is mainly derived from P $p_z$ orbitals, which is similar to the B $p_z$ orbital in MgB$_2$. One of the cylindrical Fermi surfaces in LaP$_2$H$_2$ arises largely from a hybridized state of La $d$ ($d_{xz}$ and $d_{z^2}$)/$f$ ($f_{xyz}$, $f_{xz^2}$, and $f_{x(x^2-3y^2)}$) and P $s$/$p$ orbitals [Figs. 3(b), S8, and S9], while the other is dominated by a mixed state of La $d$ ($d_{xy}$ and $d_{x^2-y^2}$)/$f$ ($f_{z^3}$ and $f_{z(x^2-y^2)}$) and P $s$/$p$ ($p_x$ and $p_y$) orbitals [Figs. 3(c), S8, and S9], which is different from the contribution of the B $s$/$p$ ($p_x$ and $p_y$) orbitals in MgB$_2$ [16].

The unique crystal structure and electronic properties encourage us to explore the superconductivity of LaP$_2$H$_2$. The calculated EPC parameter ($\lambda$) of LaP$_2$H$_2$ is 0.90 at 125 GPa [Fig. 3(e)], where the vibrations of pure P and H atoms contribute 31% and 16% to the total $\lambda$ in the range 9–18 and 29–48 THz, respectively; and the coupling between P and La atoms (below 9 THz) and P and H atoms (18–29 THz) contribute 32% and 21%, respectively. The origin of superconductivity in LaP$_2$H$_2$ is different from MgB$_2$ or H$_3$S. The role of the graphenelike P is similar to the B layer in MgB$_2$, whereas the La-P coupling becomes much stronger than Mg-B and even comparable to graphenelike P. On the other hand, the contribution of high-frequency H vibrations to $\lambda$ (16%) is much smaller than 82.6% in H$_3$S [7]. Considering that the Allen-Dynes modified McMillan equation gives a better description of superconductors with $\lambda < 1.5$ [50], we applied it to calculate the $T_c$ of LaP$_2$H$_2$. The resulting $T_c$ is 36.3 K at 125 GPa with a typical Coulomb pseudopotential parameter of $\mu^* = 0.1$, which is much higher than that of LaH$_4$ (5 K at 300 GPa) [5], YS$_4$H$_4$ (20 K at 200 GPa) [67], LaSH$_6$ (35 K at 300 GPa) [68], MgH$_2$ (23 K at 180 GPa) [69], and BH (21 K at 175 GPa) [63], with a higher H content than LaP$_2$H$_2$. This indicates that the presence of novel building units in compounds might be also an effective routine to boost the $T_c$ at a relatively lower pressure.

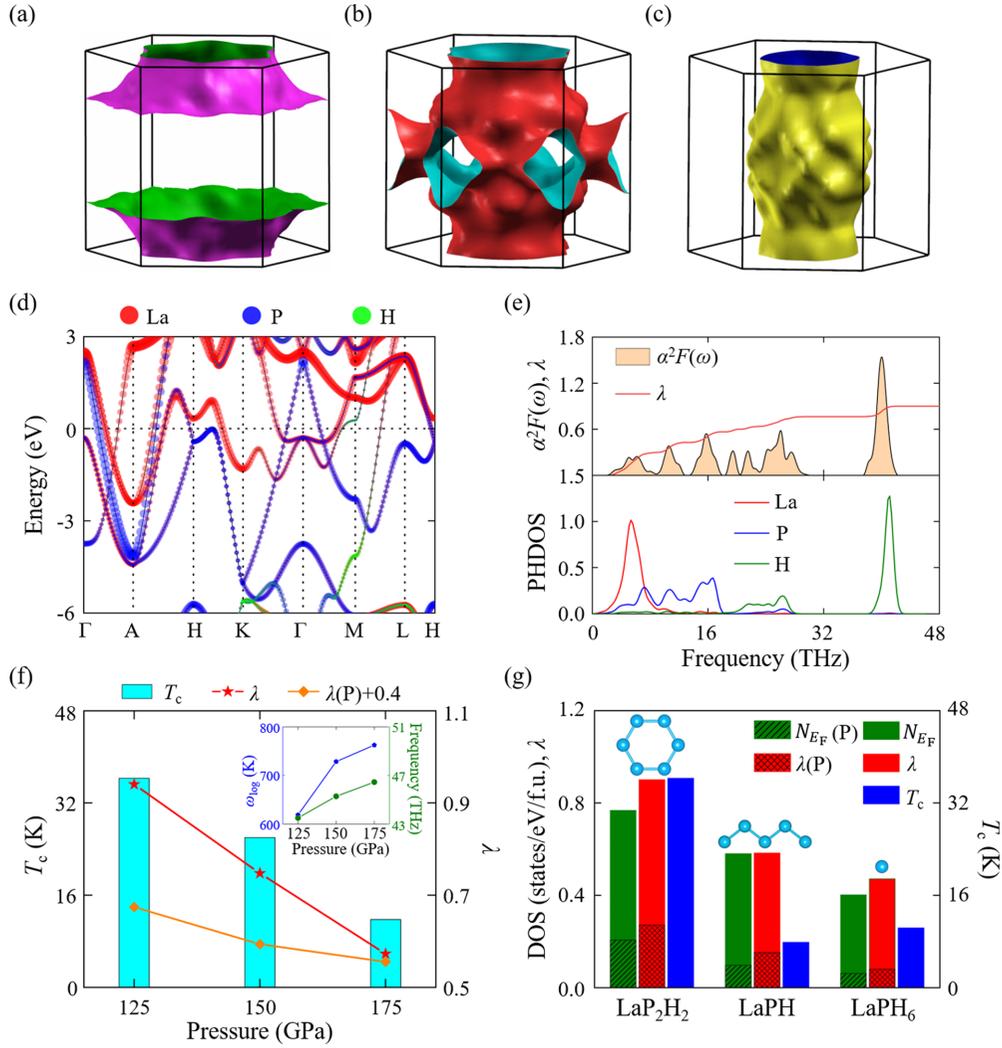

FIG. 3. (a)–(c) The three considered Fermi surfaces (the rest are in Fig. S6), (d) the projected electronic band structure, (e) the projected phonon density of states and Eliashberg spectral function of $P6/mmm$ LaP$_2$H$_2$ at 125 GPa. (f) Pressure dependence of $T_c$, the total EPC parameter $\lambda$ and the contribution of P atoms to the total $\lambda$, namely $\lambda$(P), for LaP$_2$H$_2$. The inset shows the evolution of the logarithmic average phonon frequency $\omega_{\log}$ and the highest phonon frequency of H with pressure. (g) Calculated $T_c$ values and related parameters for LaP$_2$H$_2$ at 125 GPa, LaPH at 100 GPa, and LaPH$_6$ at 200 GPa. Relative contribution of P atoms to $\lambda$, $\lambda$(P), and the DOS at the $E_F$ projected on P atomic orbitals, $N_{E_F}$(P).

Furthermore, we also analyzed the pressure-dependent superconductivity of LaP$_2$H$_2$ [Fig. 3(f)]. $T_c$ decreases with pressure, showing the same trend as $\lambda$. In contrast, $\omega_{\log}$ monotonously increases, which can be attributed to the fact that the P-H bond is easier to shorten than the P-P bond (Fig. S10), resulting in a higher frequency of H vibration.

Therefore, $\lambda$ plays a decisive role in the $T_c$ of LaP$_2$H$_2$. On the other hand, the P-P coupling always accounts for a large proportion of the total $\lambda$, and it is positively correlated with the total $\lambda$ under different pressures [Fig. 3(f)]. Consequently, we can conclude that the contribution of the rigid P layer to superconductivity does not show a strong dependence with pressure.

For LaPH, with the lowest H content, its metallicity is dominated by La atoms in the puckered layers (Fig. S5d). For LaPH$_6$, with the highest content of H, the contribution of H atoms to the DOS at the $E_F$ is the highest, but there is a clear difference between the contributions of H$_1$ and H$_2$ atoms (Fig. S5e). H$_1$ atoms show a strong contribution near the high symmetry point T, while H$_2$ atoms mainly contribute along the path U-R. On the other hand, phonon frequencies associated to H$_1$ atoms are higher than those of H$_2$ atoms (Fig. S11f), mainly due to the stronger polar covalent bond between H$_1$ and P atoms. The resulting $\lambda$ of LaPH is 0.59 at 100 GPa, where the low-frequency vibrations of La atoms (below 8 THz) contribute 51% of the total $\lambda$. The $\lambda$ of LaPH$_6$ is 0.47 at 200 GPa, mainly coming from the high-frequency vibrations of H atoms (above 20 THz, 63% of $\lambda$), especially of the H$_1$ atoms. The corresponding $T_c$ values of LaPH and LaPH$_6$ are 8.0 and 10.4 K (Figs. S11d and S11e), respectively.

Comparing different high-$T_c$ hydrides, H atoms generally show several typical structural characteristics, such as the H cages [5,8,9], 2D planes [3,65], or 3D covalent network with nonmetals [7,70]. All the H atoms in the La-P-H system are in the atomic form, but P atoms form various motifs and make a different contribution to $T_c$ [Fig. 3(g)]. From LaP$_2$H$_2$, with graphenelike P, to LaPH, with zigzag P chains, passing by LaPH$_6$, with isolated P atoms, basically, the corresponding $T_c$ decreases, which can be connected to the reduction of $\lambda$ and the DOS at the $E_F$. Additionally, the graphenelike P configuration in LaP$_2$H$_2$ makes the EPC stronger than those in LaPH and LaPH$_6$ due to the enhancement of the coupling between La and P (Fig. S11). On the whole, the discovery of LaP$_2$H$_2$ with a graphenelike P and the highest $T_c$ in La-P-H system promotes the correlation between new building blocks and novel superconductors.

In summary, our extensive first-principles structural searches for the La-P-H system under high pressure have identified three stable stoichiometries: $P6/mmm$ LaP$_2$H$_2$, $Pnma$ LaPH, and $Pmmn$ LaPH$_6$, where P atoms appear forming different building blocks. In LaP$_2$H$_2$, P atoms form graphenelike layers, in LaPH zigzag chains, and in LaPH$_6$ isolated P atoms are present. LaP$_2$H$_2$ is predicted to be a superconductor with an estimated $T_c$ of 36.3 K at 125 GPa, which is much higher than the $T_c$ predicted for LaPH$_6$, 10.4 K at 200 GPa. The graphenelike P layer and its coupling with La atoms dominate the superconducting origin of LaP$_2$H$_2$. On the other hand, LaP$_2$H$_2$ can achieve a superconducting transition at lower pressures compared with hydrides with a higher hydrogen content. The superconductivity in LaP$_2$H$_2$, LaPH, and LaPH$_6$ closely correlates with P motifs. This work will stimulate further experimental and theoretical studies to explore conventional superconductors with different building blocks.

The authors acknowledge funding from the Natural Science Foundation of China under Grants No. 21873017, No. 52022089, and No. 21573037, Natural Science Foundation of Hebei Province Grants No. A2019203507 and No. B2021203030, the Postdoctoral Science Foundation of China under Grant No. 2013M541283, and the Natural Science Foundation of Jilin Province (Grant No. 20190201231JC). This work was carried out at National Supercomputer Center in Tianjin, and the calculations were performed on TianHe-1 (A). A.B. acknowledges financial support from the Spanish Ministry of Science and Innovation (Grant No. PID2019-105488GB-I00).